\documentclass[aps,preprint]{revtex4}
\usepackage{amssymb}
\usepackage{amsmath}

\input{tcilatex}

\begin{document}

\title{Zero-norm states and High-energy Symmetries of String Theory}
\author{Chuan-Tsung Chan}
\email{ctchan@ep.nctu.edu.tw}
\affiliation{Department of International Trade, Lan Yang Institute of Technology, I-Lan,
Taiwan, R.O.C.}
\author{Jen-Chi Lee}
\email{jcclee@cc.nctu.edu.tw}
\affiliation{Department of Electrophysics, National Chiao-Tung University, Hsinchu,
Taiwan, R.O.C.}
\date{\today }

\begin{abstract}
We derive stringy Ward identities from the decoupling of two types of
zero-norm states in the old covariant first quantized (OCFQ) spectrum of
open bosonic string. These Ward identities are valid to all energy $\alpha
^{\prime }$ and all loop orders $\chi $ in string perturbation theory. The
high-energy limit $\alpha ^{\prime }\rightarrow \infty $ of these stringy
Ward identities can then be used to fix the proportionality constants
between scattering amplitudes of different string states algebraically 
\textit{without} referring to Gross and Mende's saddle point calculation of
high-energy string-loop amplitudes. As examples, all Ward identities for the
mass level $M^{2}$ $=4,6$ are derived, their high-energy limits are
calculated and the the proportionality constants between scattering
amplitudes of different string states are determined. In addition to those
identified before, we discover some \textit{new} nonzero components of
high-energy amplitudes not found previously by Gross and Manes. These
components are essential to preserve massive gauge invariances or decouple
massive zero-norm states of string theory. A set of massive scattering
amplitudes and their high energy limits are calculated explicitly for each
mass level $M^{2}$ $=4,6$ to justify our results.
\end{abstract}

\maketitle

\section{Introduction}

It is often of fundamental importance to study the high-energy behavior of a
local quantum field theory. In the quantum chromodynamics, for example, the
renormalization group and the discovery of asymptotic freedom \cite{1}
turned out to be one of the most important properties of Yang-Mills
theories. On the other hand, the spontaneously broken symmetries are often
hidden at low energy, but become evident in the high-energy behavior of the
theory. In string theory, one expects even more rich fundamental structures
at high-energy since only then will an infinite number of particles be
excited. Being a consistent quantum theory with no free parameter, it is
conceivable that an huge symmetry group or Ward identities get restored at
high-energy, which are responsible for the ultraviolet finiteness of string
theory.

Recently it was discovered that \cite{2} the high-energy limits $\alpha
^{\prime }\rightarrow \infty $ of stringy Ward identities can be used to fix
the proportionality constants between scattering amplitudes of different
string states algebraically \textit{without} referring to Gross and Mende's 
\cite{3} saddle point calculation of high-energy string-loop amplitudes.
These proportionality constants are, as conjectured by Gross \cite{4},
independent of the scattering angle $\phi _{CM}$ and the order $\chi $ of
string perturbation theory. As a result, all high-energy string scattering
amplitudes can be expressed in terms of those of tachyons. These Ward
identities, which are valid to all energy $\alpha ^{\prime }$ and all loop
orders $\chi $ in string perturbation theory, are derived from the
decoupling of two types of zero-norm states in the old covariant first
quantized (OCFQ) spectrum of open bosonic string. A prescription to
explicitly calculate zero-norm states for arbitrary mass levels, or stringy
symmetry charges with arbitrarily high spins, was given in \cite{5}. The
importance of zero-norm states and their implication on stringy symmetries
were first pointed out in the context of massive $\sigma $-model approach of
string theory \cite{6}. These stringy symmetries were also demonstrated
recently in Witten's string field theory (WSFT), and the background ghost
fields in the off-shell BRST spectrum were identified, in a one to one
manner, to the lifting of the on-shell conditions of zero-norm states in the
OCFQ approach \cite{7}. On the other hand, zero-norm states were also shown 
\cite{8} to carry the spacetime $\omega _{\infty }$ symmetry charges of toy
2D string theory, and the corresponding $\omega _{\infty }$ Ward identities
were powerful enough to determine the tachyon scattering amplitudes
algebraically \textit{without }any integration \cite{9}.

In this paper, all Ward identities for the mass level $M^{2}$ $=4,6$ will be
derived, their high-energy limits are calculated and the the proportionality
constants between scattering amplitudes of different string states are
determined directly from these Ward identities. General formula of
high-energy amplitudes for arbitrary mass levels will be given in terms of
those of tachyons. \textit{In addition to those identified before, we
discover some new nonzero components of high-energy amplitudes at each mass
level not found previously by Gross and Manes \cite{10}. These components
are essential to preserve massive gauge invariances or decouple massive
zero-norm states of string theory.} A set of massive scattering amplitudes
and their high energy limits are calculated explicitly for each mass level $%
M^{2}$ $=4,6$ to justify our results. This paper is organized as following.
In section II, we derive stringy Ward identities for the mass level $M^{2}$ $%
=4$ \cite{11}, and then take high-energy limits of them to determine the
proportionality constants between scattering amplitudes of different string
states algebraically. At the subleading order energy, one finds 6 unknown
amplitudes and 4 equations. Presumably, they are not proportional to each
other or the proportional coefficients do depend on the scattering angle $%
\phi _{CM}$. This result will be confirmed at section III. In section III,
the high energy limits of a set of string-tree level amplitudes with one
tensor at mass level $M^{2}$ $=4$ and three tachyons are explicitly
calculated to justify the results of section II. The whole program is then
generalized to mass level $M^{2}$ $=6$ in section IV. We make a comparision
of our results with those of Gross and Manes \cite{10} in section V. Finally
a brief conclusion is given in section VI.

\section{High-energy stringy Ward identities of mass level $M^{2}$ $=4$}

In the OCFQ spectrum of open bosonic string theory, the solutions of
physical states conditions include positive-norm propagating states and two
types of zero-norm states which were neglected in the most literature. They
are \cite{12}

\begin{equation}
\text{Type I}:L_{-1}\left\vert x\right\rangle ,\text{ where }L_{1}\left\vert
x\right\rangle =L_{2}\left\vert x\right\rangle =0,\text{ }L_{0}\left\vert
x\right\rangle =0;  \tag{2.1}
\end{equation}

\begin{equation}
\text{Type II}:(L_{-2}+\frac{3}{2}L_{-1}^{2})\left\vert \widetilde{x}%
\right\rangle ,\text{ where }L_{1}\left\vert \widetilde{x}\right\rangle
=L_{2}\left\vert \widetilde{x}\right\rangle =0,\text{ }(L_{0}+1)\left\vert 
\widetilde{x}\right\rangle =0.  \tag{2.2}
\end{equation}%
Equations (2.1) and (2.2) can be derived from Kac determinant in conformal
field theory. While type I states have zero-norm at any spacetime dimension,
type II states have zero-norm \textit{only} at D=26. The existence of type
II zero-norm states signals the importance of zero-norm states in the
structure of the theory of string. In the first quantized approach of string
theory, the stringy \textit{on-shell} Ward identities are proposed to be
(for our purpose we choose four-point amplitudes in this paper)

\begin{equation}
\mathcal{T}_{\chi }(k_{i})=g^{2-\chi }\int \frac{Dg_{\alpha \beta }}{%
\mathcal{N}}DX^{\mu }\exp (-\frac{\alpha ^{\prime }}{2\pi }\int d^{2}\xi 
\sqrt{g}g^{\alpha \beta }\partial _{\alpha }X^{\mu }\partial _{\beta }X_{\mu
})\overset{4}{\underset{i=1}{\Pi }}v_{i}(k_{i})=0,  \tag{2.3}
\end{equation}%
where at least one of the 4 vertex operators corresponds to the zero-norm
state solution of eqs. (2.1) or (2.2). In eq(2.3) $g$ is the string coupling
constant, $\mathcal{N}$ is the volume of the group of diffeomorphisms and
Weyl rescalings of the worldsheet metric, and $v_{i}(k_{i})$ are the
on-shell vertex operators with momenta $k_{i}$. The integral is over
orientable open surfaces of Euler number $\chi $ parametrized by moduli $%
\overrightarrow{m}$ with punctures at $\xi _{i}$. The simplest zero-norm
state $k\cdot \alpha _{-1}\mid 0,k\rangle $, $%
k^{2}=0%
$ with polarization $k$ is the massless solution of eq. (2.1), which
reproduces the Ward identity of string QED when substituting into eq(2.3). A
simple \ prescription to systematically solve eqs. (2.1) and (2.2) for an
infinite number of zero-norm states was given in \cite{5}. A more thorough
understanding of the solution of these equations and their relation to
space-time $\omega _{\infty }$ symmetry of toy D=2 string was discussed in 
\cite{8}. For our purpose here, there are four zero-norm states at mass
level \ $M^{2}$ $=4$, the corresponding Ward identities were calculated to
be \cite{11}

\begin{equation}
k_{\mu }\theta _{\nu \lambda }\mathcal{T}_{\chi }^{(\mu \nu \lambda
)}+2\theta _{\mu \nu }\mathcal{T}_{\chi }^{(\mu \nu )}=0,  \tag{2.4}
\end{equation}%
\begin{equation}
(\frac{5}{2}k_{\mu }k_{\nu }\theta _{\lambda }^{\prime }+\eta _{\mu \nu
}\theta _{\lambda }^{\prime })\mathcal{T}_{\chi }^{(\mu \nu \lambda
)}+9k_{\mu }\theta _{\nu }^{\prime }\mathcal{T}_{\chi }^{(\mu \nu )}+6\theta
_{\mu }^{\prime }\mathcal{T}_{\chi }^{\mu }=0,  \tag{2.5}
\end{equation}%
\begin{equation}
(\frac{1}{2}k_{\mu }k_{\nu }\theta _{\lambda }+2\eta _{\mu \nu }\theta
_{\lambda })\mathcal{T}_{\chi }^{(\mu \nu \lambda )}+9k_{\mu }\theta _{\nu }%
\mathcal{T}_{\chi }^{[\mu \nu ]}-6\theta _{\mu }\mathcal{T}_{\chi }^{\mu }=0,
\tag{2.6}
\end{equation}%
\begin{equation}
(\frac{17}{4}k_{\mu }k_{\nu }k_{\lambda }+\frac{9}{2}\eta _{\mu \nu
}k_{\lambda })\mathcal{T}_{\chi }^{(\mu \nu \lambda )}+(9\eta _{\mu \nu
}+21k_{\mu }k_{\nu })\mathcal{T}_{\chi }^{(\mu \nu )}+25k_{\mu }\mathcal{T}%
_{\chi }^{\mu }=0,  \tag{2.7}
\end{equation}%
where $\theta _{\mu \nu }$ is transverse and traceless, and $\theta
_{\lambda }^{\prime }$ and $\theta _{\lambda }$ are transverse vectors. In
each equation, we have chosen, say, $v_{2}(k_{2})$\ to be the vertex
operators constructed from zero-norm states and $k_{\mu }\equiv k_{2\mu }$.
Note that eq.(2.6) is the inter-particle Ward identity corresponding to $%
D_{2}$ vector zero-norm state obtained by antisymmetrizing those terms which
contain $\alpha _{-1}^{\mu }\alpha _{-2}^{\nu }$ in the original type I and
type II vector zero-norm states. We will use 1 and 2 for the incoming
particles and 3 and 4 for the scattered particles. In eqs. (2.4)-(2.7), 1,3
and 4 can be any string states (including zero-norm states) and we have
omitted their tensor indices for the cases of excited string states. For
example, one can choose $v_{1}(k_{1})$\ to be the vertex operator
constructed from another zero-norm state which generates an inter-particle
Ward identity of the third massive level. The resulting Ward-identity of eq
(2.6) then relates scattering amplitudes of particles at different mass
level. $\mathcal{T}_{\chi }^{\prime }s$ in eqs (2.4)-(2.7) are the mass
level $M^{2}$ $=4$, $\chi $-th order string-loop amplitudes. At this point,
\{$\mathcal{T}_{\chi }^{(\mu \nu \lambda )},\mathcal{T}_{\chi }^{(\mu \nu )},%
\mathcal{T}_{\chi }^{\mu }$\} is identified to be the \emph{amplitude triplet%
} of the spin-three state. $\mathcal{T}_{\chi }^{[\mu \nu ]}$ is obviously
identified to be the scattering amplitude of the antisymmetric spin-two
state with the same momenta as $\mathcal{T}_{\chi }^{(\mu \nu \lambda )}$.
Eq. (2.6) thus relates the scattering amplitudes of two different string
states at mass level $M^{2}$ $=4$. Note that eqs. (2.4)-(2.7) are valid
order by order and are \emph{automatically} of the identical form in string
perturbation theory. This is consistent with Gross's argument through the
calculation of high-energy scattering amplitudes. However, it is important
to note that eqs. (2.4)-(2.7) are, in contrast to the high-energy $\alpha
^{\prime }\rightarrow \infty $ result of Gross, valid to \emph{all} energy $%
\alpha ^{\prime }$ and their coefficients do depend on the center of mass
scattering angle $\phi _{CM}$ , which is defined to be the angle between $%
\overrightarrow{k}_{1}$ and $\overrightarrow{k}_{3}$, through the dependence
of momentum $k$ .

We will calculate high energy limit of eqs.(2.4)-(2.7) without referring to
the saddle point calculation in \cite{3,4,10}. Let's define the normalized
polarization vectors

\begin{equation}
e_{P}=\frac{1}{m_{2}}(E_{2},\mathrm{k}_{2},0)=\frac{k_{2}}{m_{2}},  \tag{2.8}
\end{equation}

\begin{equation}
e_{L}=\frac{1}{m_{2}}(\mathrm{k}_{2},E_{2},0),  \tag{2.9}
\end{equation}%
\begin{equation}
e_{T}=(0,0,1)  \tag{2.10}
\end{equation}%
in the CM frame contained in the plane of scattering. They satisfy the
completeness relation

\begin{equation}
\eta ^{\mu \nu }=\underset{\alpha ,\beta }{\tsum }e_{\alpha }^{\mu }e_{\beta
}^{\nu }\eta ^{\alpha \beta }  \tag{2.11}
\end{equation}%
where $\mu ,\nu =0,1,2$ and $\alpha ,\beta =P,L,T.$ $Diag$ $\eta ^{\mu \nu
}=(-1,1,1).$One can now transform all $\mu ,\nu $ coordinates in
eqs.(2.4)-(2.7) to coordinates $\alpha ,\beta $. For eq(2.4), we have $%
\theta ^{\mu \nu }=e_{L}^{\mu }e_{L}^{\nu }-e_{T}^{\mu }e_{T}^{\nu }$ or $%
\theta ^{\mu \nu }=e_{L}^{\mu }e_{T}^{\nu }+e_{T}^{\mu }e_{L}^{\nu }$ . In
the high energy $E\rightarrow $ $\infty ,$ fixed angle $\phi _{CM}$ limit,
one identifies $e_{P}=e_{L}$ and eq. (2.4) gives ( we drop loop order $\chi $
here to simplify the notation)

\begin{equation}
\mathcal{T}_{LLL}^{6\rightarrow 4}-\mathcal{T}_{LTT}^{4}+\mathcal{T}%
_{(LL)}^{4}-\mathcal{T}_{(TT)}^{2}=0,  \tag{2.12}
\end{equation}%
\begin{equation}
\mathcal{T}_{LLT}^{5\rightarrow 3}+\mathcal{T}_{(LT)}^{3}=0.  \tag{2.13}
\end{equation}%
In eqs (2.12) and (2.13), we have assigned a relative energy power for each
amplitude. For each longitudinal $L$ component, the order is $E^{2}$and for
each transverse $T$ component, the order is $E.$ This is due to the
definitions of $e_{L}$and $e_{T}$ in eqs (2.9) and (2.10), where $e_{L}$ got
one energy power more than $e_{T}.$By eq. (2.12), the $E^{6}$ term of the
energy expansion for $\mathcal{T}_{LLL}$ is forced to be zero. As a result,
the possible leading order term is $E^{4}$. Similar rule applies to $%
\mathcal{T}_{LLT}$ in eq(2.13). For eq(2.5), we have $\theta ^{\prime \mu
}=e_{L}^{\mu }$ or $\theta ^{\prime \mu }=e_{T}^{\mu }$ and one gets, in the
high energy limit,

\begin{equation}
10\mathcal{T}_{LLL}^{6\rightarrow 4}+\mathcal{T}_{LTT}^{4}+18\mathcal{T}%
_{(LL)}^{4}+6\mathcal{T}_{L}^{2}=0,  \tag{2.14}
\end{equation}%
\begin{equation}
10\mathcal{T}_{LLT}^{5\rightarrow 3}+\mathcal{T}_{TTT}^{3}+18\mathcal{T}%
_{(LT)}^{3}+6\mathcal{T}_{T}^{1}=0.  \tag{2.15}
\end{equation}%
For the $D_{2}$ Ward identity, eq.(2.6), we have $\theta ^{\mu }=e_{L}^{\mu
} $ or $\theta ^{\mu }=e_{T}^{\mu }$ and one gets, in the high energy limit,

\begin{equation}
\mathcal{T}_{LLL}^{6\rightarrow 4}+\mathcal{T}_{LTT}^{4}+9\mathcal{T}%
_{[LL]}^{4\rightarrow 2}-3\mathcal{T}_{L}^{2}=0,  \tag{2.16}
\end{equation}

\begin{equation}
\mathcal{T}_{LLT}^{5\rightarrow 3}+\mathcal{T}_{TTT}^{3}+9\mathcal{T}%
_{[LT]}^{3}-3\mathcal{T}_{T}^{1}=0.  \tag{2.17}
\end{equation}%
It is important to note that $\mathcal{T}_{[LL]}$ in eq.(2.16) originate
from the high energy limit of $\mathcal{T}_{[PL]}$, and the antisymmetric
property of the tensor forces the leading $E^{4}$ term to be zero. Finally
the singlet zero norm state Ward identity, eq.(2.7), imply, in the high
energy limit,

\begin{equation}
34\mathcal{T}_{LLL}^{6\rightarrow 4}+9\mathcal{T}_{LTT}^{4}+84\mathcal{T}%
_{(LL)}^{4}+9\mathcal{T}_{(TT)}^{2}+50\mathcal{T}_{L}^{2}=0.  \tag{2.18}
\end{equation}%
One notes that all components of high energy amplitudes of symmetric spin
three and antisymmetric spin two states appear at least once in eqs.
(2.12)-(2.18). It is now easy to see that the naive leading order amplitudes
corresponding to $E^{4}$ \ appear in eqs.(2.12), (2.14), (2.16) and (2.18).
However, a simple calculation shows that $\mathcal{T}_{LLL}^{4}=\mathcal{T}%
_{LTT}^{4}=\mathcal{T}_{(LL)}^{4}=0.$So the real leading order amplitudes
correspond to $E^{3}$, which appear in eqs.(2.13), (2.15) and (2.17). A
simple calculation shows that

\begin{equation}
\mathcal{T}_{TTT}^{3}:\mathcal{T}_{LLT}^{3}:\mathcal{T}_{(LT)}^{3}:\mathcal{T%
}_{[LT]}^{3}=8:1:-1:-1.  \tag{2.19}
\end{equation}%
Note that these proportionality constants are, as conjectured by Gross \cite%
{4}, independent of the scattering angle $\phi _{CM}$ and the loop order $%
\chi $ of string perturbation theory. They are also independent of particles
chosen for vertex $v_{1,3,4}$. \textit{Most importantly, we now understand
that they originate from zero-norm states in the OCFQ spectrum of the
string! }The subleading order amplitudes corresponding to $E^{2}$appear in
eqs.(2.12), (2.14), (2.16) and (2.18). One has 6 unknown amplitudes and 4
equations. Presumably, they are not proportional to each other or the
proportional coefficients do depend on the scattering angle $\phi _{CM}$. We
will justify this point later in our sample calculation in section III. Our
calculation here is purely algebraic \textit{without any integration} and is
independent of saddle point calculation in \cite{3,4,10}. It is important to
note that our result in eq.(2.19) is gauge invariant as it should be since
we derive it from Ward identities (2.4)-(2.7). On the other hand, the result
obtained in \cite{10} with $\mathcal{T}_{TTT}^{3}\propto \mathcal{T}%
_{[LT]}^{3},$ and $\mathcal{T}_{LLT}^{3}=0$ in the leading order energy at
this mass level is, on the contrary, \textit{not} gauge invariant. In fact,
with $\mathcal{T}_{LLT}^{3}=0$, an inconsistency arises, for example,
between eqs. (2.13) and (2.15). We give one example here to illustrate the
meaning of the massive gauge invariant amplitude. To be more specific, we
will use two different gauge choices to calculate the high-energy scattering
amplitude of symmetric spin three state. The first gauge choice is

\begin{equation}
(\epsilon _{\mu \nu \lambda }\alpha _{-1}^{\mu \nu \lambda }+\epsilon _{(\mu
\nu )}\alpha _{-1}^{\mu }\alpha _{-2}^{\nu })\left\vert 0,k\right\rangle
;\epsilon _{(\mu \nu )}=-\frac{3}{2}k^{\lambda }\epsilon _{\mu \nu \lambda
},k^{\mu }k^{\nu }\epsilon _{\mu \nu \lambda }=0,\eta ^{\mu \nu }\epsilon
_{\mu \nu \lambda }=0.  \tag{2.20}
\end{equation}%
In the high-energy limit, using the helixity decomposition and writing $%
\epsilon _{\mu \nu \lambda }=\Sigma _{\mu ,\nu ,\lambda }e_{\mu }^{\alpha
}e_{\nu }^{\beta }e_{\lambda }^{\delta }u_{\alpha \beta \delta };\alpha
,\beta ,\delta =P,L,T,$ we get

\begin{eqnarray}
(\epsilon _{\mu \nu \lambda }\alpha _{-1}^{\mu \nu \lambda }+\epsilon _{(\mu
\nu )}\alpha _{-1}^{\mu }\alpha _{-2}^{\nu })\left\vert 0,k\right\rangle
&=&[u_{PLT}(6\alpha _{-1}^{PLT}+6\alpha _{-1}^{(L}\alpha _{-2}^{T)})  \notag
\\
&&+u_{TTP}(3\alpha _{-1}^{TTP}-3\alpha _{-1}^{LLP}+3\alpha _{-1}^{(T}\alpha
_{-2}^{T)}-3\alpha _{-1}^{(L}\alpha _{-2}^{L)})  \notag \\
&&+u_{TTL}(3\alpha _{-1}^{TTL}-\alpha _{-1}^{LLL})+u_{TTT}(\alpha
_{-1}^{TTT}-3\alpha _{-1}^{LLT})]\left\vert 0,k\right\rangle .  \TCItag{2.21}
\end{eqnarray}%
The second gauge choice is

\begin{equation}
\widetilde{\varepsilon }_{\mu \nu \lambda }\alpha _{-1}^{\mu \nu \lambda
}\left\vert 0,k\right\rangle ;k^{\mu }\widetilde{\varepsilon }_{\mu \nu
\lambda }=0,\eta ^{\mu \nu }\widetilde{\varepsilon }_{\mu \nu \lambda }=0. 
\tag{2.22}
\end{equation}%
In the high-energy limit, similar calculation gives

\begin{equation}
\widetilde{\varepsilon }_{\mu \nu \lambda }\alpha _{-1}^{\mu \nu \lambda
}\left\vert 0,k\right\rangle =[\widetilde{u}_{TTL}(3\alpha
_{-1}^{TTL}-\alpha _{-1}^{LLL})+\widetilde{u}_{TTT}(\alpha
_{-1}^{TTT}-3\alpha _{-1}^{LLT})]\left\vert 0,k\right\rangle .  \tag{2.23}
\end{equation}

It is now easy to see that the first and second terms of eq.(2.21) will not
contribute to the high-energy scattering amplitude of the symmetric spin
three state due to the spin two Ward identities eqs.(2.13) and (2.12) if we
identify $e_{P}=e_{L}.$ Thus the two different gauge choices eqs.(2.20) and
(2.22) give the same high-energy scattering amplitude. It can be shown that
this massive gauge symmetry is valid to all energy and is the result of the
decoupling of massive spin two zero-norm state at mass level $M^{2}$ $=4.$
Note that the $\alpha _{-1}^{LLT}$ term of eq.(2.23), which corresponds to
the amplitude $\mathcal{T}_{LLT}^{3},$ was missing in the calculation of Ref 
\cite{10}. We will discuss this issue in section V.

To further justify our result, we give a sample calculation in section III.

\section{A sample calculation of mass level $M^{2}$ $=4$}

In this section, we give a detailed calculation of a set of sample
scattering amplitudes to explicitly justify our results presented in section
II. Since the proportionality constants in eq.(2.19) are independent of
particles chosen for vertex $v_{1,3,4}.$ For simplicity, we will choose them
to be tachyons. For the string-tree level $\chi =1$, with one tensor $v_{2}$
and three tachyons $v_{1,3,4}$, all scattering amplitudes of mass level $%
M^{2}$ $=4$ were calculated in \cite{11}. They are ( $s-t$ channel only)

\begin{eqnarray}
\mathcal{T}^{\mu \nu \lambda } &=&\tint
\tprod_{i=1}^{4}dx_{i}<e^{ik_{1}X}\partial X^{\mu }\partial X^{\nu }\partial
X^{\lambda }e^{ik_{2}X}e^{ik_{3}X}e^{ik_{4}X}>  \notag \\
&=&\frac{\Gamma (-\frac{s}{2}-1)\Gamma (-\frac{t}{2}-1)}{\Gamma (\frac{u}{2}%
+2)}[{-{t/2}({t^{2}}/4-1)k_{1}^{\mu }k_{1}^{\nu }k_{1}^{\lambda
}+3(s/2+1)t/2(t/2+1)k_{1}^{(\mu }k_{1}^{\nu }k_{3}^{\lambda )}}  \notag \\
&&{-3s/2(s/2+1)(t/2+1)k_{1}^{(\mu }k_{3}^{\nu }k_{3}^{\lambda )}+s/2({s^{2}}%
/4-1)k_{3}^{\mu }k_{3}^{\nu }k_{3}^{\lambda }]},  \TCItag{3.1}
\end{eqnarray}

\begin{eqnarray}
\mathcal{T}^{(\mu \nu )} &=&\tint \tprod_{i=1}^{4}dx_{i}<e^{ik_{1}X}\partial
^{2}X^{(\mu }\partial X^{\nu )}e^{ik_{2}X}e^{ik_{3}X}e^{ik_{4}X}>  \notag \\
&=&\frac{\Gamma (-\frac{s}{2}-1)\Gamma (-\frac{t}{2}-1)}{\Gamma (\frac{u}{2}%
+2)}[{t/2({t^{2}}/4-1)k_{1}^{\mu }k_{1}^{\nu }-(s/2+1)t/2(t/2+1)k_{1}^{(\mu
}k_{3}^{\nu )}}  \notag \\
&&{+\newline
s/2(s/2+1)(t/2+1)k_{3}^{(\mu }k_{1}^{\nu )}-s/2({s^{2}}/4-1)k_{3}^{\mu
}k_{3}^{\nu }]},  \TCItag{3.2}
\end{eqnarray}

\begin{eqnarray}
\mathcal{T}^{\mu } &=&\frac{1}{2}\tint
\tprod_{i=1}^{4}dx_{i}<e^{ik_{1}X}\partial ^{3}X^{\mu
}e^{ik_{2}X}e^{ik_{3}X}e^{ik_{4}X}>  \notag \\
&=&\frac{\Gamma (-\frac{s}{2}-1)\Gamma (-\frac{t}{2}-1)}{\Gamma (\frac{u}{2}%
+2)}[s/2({s^{2}}/4-1)k_{3}^{\mu }-t/2({t^{2}}/4-1)k_{1}^{\mu }], 
\TCItag{3.3}
\end{eqnarray}

\begin{eqnarray}
\mathcal{T}^{[\mu \nu ]} &=&\tint \tprod_{i=1}^{4}dx_{i}<e^{ik_{1}X}\partial
^{2}X^{[\mu }\partial X^{\nu ]}e^{ik_{2}X}e^{ik_{3}X}e^{ik_{4}X}>  \notag \\
&=&\frac{\Gamma (-\frac{s}{2}-1)\Gamma (-\frac{t}{2}-1)}{\Gamma (\frac{u}{2}%
+2)}[(\frac{s+t}{2})(s/2+1)(t/2+1)k_{3}^{[\mu }k_{1}^{\nu ]}]  \TCItag{3.4}
\end{eqnarray}%
where $%
s=-(k_{1}+k_{2})^{2}%
$, $%
t=-(k_{2}+k_{3})^{2}%
$and $%
u=-(k_{1}+k_{3})^{2}%
$ are the Mandelstam variables. In deriving eqs. (3.1) to (3.4), we have
made the $SL(2,R)$ gauge fixing by choosing $x_{1}=0,0\leqq x_{2}\leqq
1,x_{3}=1,x_{4}=\infty .$ To calculate the high energy expansions $%
(s,t\rightarrow \infty ,\frac{s}{t}=$ fixed $)$ of these scattering
amplitudes, one needs the followong energy expansion formulas 
\begin{equation}
e_{P}.k_{1}=(\frac{-2E^{2}}{m_{2}})[1-(\frac{m_{2}^{2}-2}{4})\frac{1}{E^{2}}%
],  \tag{3.5}
\end{equation}

\begin{equation}
e_{L}.k_{1}=(\frac{-2E^{2}}{m_{2}})[1-(\frac{m_{2}^{2}-2}{4})\frac{1}{E^{2}}%
+(\frac{m_{2}^{2}}{4})\frac{1}{E^{4}}+(\frac{m_{2}^{4}-2m_{2}^{2}}{16})\frac{%
1}{E^{6}}+O(\frac{1}{E^{8}})],  \tag{3.6}
\end{equation}

\begin{equation}
e_{T}.k_{1}=0,  \tag{3.7}
\end{equation}

\begin{equation}
e_{P}.k_{3}=(\frac{E^{2}}{m_{2}})\left\{ 2\xi ^{2}+[\frac{m_{2}^{2}}{2}\eta
^{2}+(3\xi ^{2}-1)]\frac{1}{E^{2}}+(2\xi ^{2}-1)(\frac{m_{2}^{2}+2}{4})^{2}%
\frac{1}{E^{6}}+O(\frac{1}{E^{8}})\right\} ,  \tag{3.8}
\end{equation}

\begin{equation}
e_{L}.k_{3}=(\frac{E^{2}}{m_{2}})\left\{ 
\begin{array}{c}
2\xi ^{2}+[-\frac{m_{2}^{2}}{2}\eta ^{2}+(3\xi ^{2}-1)]\frac{1}{E^{2}}+(%
\frac{m_{2}^{2}}{2}\xi ^{2})\frac{1}{E^{4}} \\ 
+(\frac{m_{2}^{4}-4m_{2}^{2}\xi ^{2}+8\xi ^{2}-4}{16})\frac{1}{E^{6}}+O(%
\frac{1}{E^{8}})%
\end{array}%
\right\} ,  \tag{3.9}
\end{equation}

\begin{equation}
e_{T}.k_{3}=(-2\xi \eta )E-(\frac{2\xi \eta }{E})+(\frac{\xi \eta }{E^{3}})-(%
\frac{\xi \eta }{E^{5}})+O(\frac{1}{E^{7}})  \tag{3.10}
\end{equation}%
where $\xi =\sin \frac{\phi _{CM}}{2}$ and $\eta =\cos \frac{\phi _{CM}}{2}.$
The high-energy expansions of Mandelstam variables are given by

\begin{equation}
s=(E_{1}+E_{2})^{2}=4E^{2},  \tag{3.11}
\end{equation}

\begin{equation}
t=(-4\xi ^{2})E^{2}+(m_{2}^{2}-6)\xi ^{2}+\frac{1}{8}(m_{2}^{2}+2)^{2}(1-2%
\xi ^{2})\frac{1}{E^{4}}+O(\frac{1}{E^{6}}).  \tag{3.12}
\end{equation}%
We can now explicitly calculate all amplitudes in eq.(2.19). After some
algebra, we get

\begin{equation}
\mathcal{T}_{TTT}=-8E^{9}\mathcal{T}(3)\sin ^{3}\phi _{CM}[1+\frac{3}{E^{2}}+%
\frac{5}{4E^{4}}-\frac{5}{4E^{6}}+O(\frac{1}{E^{8}})],  \tag{3.13}
\end{equation}

\begin{eqnarray}
\mathcal{T}_{LLT} &=&-E^{9}\mathcal{T}(3)[\sin ^{3}\phi _{CM}+(6\sin \phi
_{CM}\cos ^{2}\phi _{CM})\frac{1}{E^{2}}  \notag \\
&&-\sin \phi _{CM}(\frac{11}{2}\sin ^{2}\phi _{CM}-6)\frac{1}{E^{4}}+O(\frac{%
1}{E^{6}})],  \TCItag{3.14}
\end{eqnarray}

\begin{eqnarray}
\mathcal{T}_{[LT]} &=&E^{9}\mathcal{T}(3)[\sin ^{3}\phi _{CM}-(2\sin \phi
_{CM}\cos ^{2}\phi _{CM})\frac{1}{E^{2}}  \notag \\
&&+\sin \phi _{CM}(\frac{3}{2}\sin ^{2}\phi _{CM}-2)\frac{1}{E^{4}}+O(\frac{1%
}{E^{6}})],  \TCItag{3.15}
\end{eqnarray}

\begin{eqnarray}
\mathcal{T}_{(LT)} &=&E^{9}\mathcal{T}(3)[\sin ^{3}\phi _{CM}+\sin \phi
_{CM}(\frac{3}{2}-10\cos \phi _{CM}  \notag \\
&&-\frac{3}{2}\cos ^{2}\phi _{CM})\frac{1}{E^{2}}-\sin \phi _{CM}(\frac{1}{4}%
+10\cos \phi _{CM}+\frac{3}{4}\cos ^{2}\phi _{CM})\frac{1}{E^{4}}+O(\frac{1}{%
E^{6}})]  \TCItag{3.16}
\end{eqnarray}%
where $\mathcal{T}(n)\mathcal{=}\sqrt{\pi }(-1)^{n-1}2^{-n}E^{-1-2n}(\sin 
\frac{\phi _{CM}}{2})^{-3}(\cos \frac{\phi _{CM}}{2})^{5-2n}\exp (-\frac{%
s\ln s+t\ln t-(s+t)\ln (s+t)}{2})$ is the high-energy limit of $\frac{\Gamma
(-\frac{s}{2}-1)\Gamma (-\frac{t}{2}-1)}{\Gamma (\frac{u}{2}+2)}$ with $%
s+t+u=2n-8$, and we have calculated it up to the next leading order in $E$.
We thus have justified eq.(2.19) with $\mathcal{T}_{TTT}^{3}=-8E^{9}\mathcal{%
T}(3)\sin ^{3}\phi _{CM}$ and $\mathcal{T}_{LLT}^{5}=0.$ We have also
checked that $\mathcal{T}_{LLL}^{6}=\mathcal{T}_{LLL}^{4}=\mathcal{T}%
_{LTT}^{4}=\mathcal{T}_{(LL)}^{4}=0$ as claimed in section II. Note that,
unlike the leading $\ E^{9}$ order, the angular dependences of $E^{7}$ order
are different for each amplitudes. The subleading order amplitudes
corresponding to $\mathcal{T}^{2}$ ($E^{8}$ order) appear in eqs.(2.12),
(2.14), (2.16) and (2.18). One has 6 unknown amplitudes. An explicit sample
calculation gives

\begin{equation}
\mathcal{T}_{LLL}^{2}=-4E^{8}\sin \phi _{CM}\cos \phi _{CM}\mathcal{T}(3), 
\tag{3.17}
\end{equation}

\begin{equation}
\mathcal{T}_{LTT}^{2}=-8E^{8}\sin ^{2}\phi _{CM}\cos \phi _{CM}\mathcal{T}%
(3),  \tag{3.18}
\end{equation}%
which show that their angular dependences are indeed different or the
proportional coefficients do depend on the scattering angle $\phi _{CM}$.

\section{The calculation of mass level $M^{2}$ $=6$}

In this section we generalize the calculation of sections II and III to mass
level $M^{2}$ $=6$. There are four positive-norm physical propagating states
at this mass level \cite{13}, a totally symmetric spin four state, a mixed
symmetric spin three state, a symmetric spin two state and a scalar state.
There are nine zero-norm states at this mass level. One can use the simplied
method \cite{5} to calculate all of them. The spin three and spin two
zero-norm states are (from now on, unless otherwise stated, each spin
polarization is assumed to be transverse, traceless and is symmetric with
respect to each group of indices)

\begin{equation}
L_{-1}\left\vert x\right\rangle =\theta _{\mu \nu \lambda }(k_{\beta }\alpha
_{-1}^{\mu \nu \lambda \beta }+3\alpha _{-1}^{\mu \nu }\alpha _{-2}^{\lambda
})\left\vert 0,k\right\rangle ;\left\vert x\right\rangle =\theta _{\mu \nu
\lambda }\alpha _{-1}^{\mu \nu \lambda }\left\vert 0,k\right\rangle , 
\tag{4.1}
\end{equation}

\begin{equation}
L_{-1}\left\vert x\right\rangle =[k_{\lambda }\theta _{\mu \nu }\alpha
_{-1}^{\mu _{\lambda }}\alpha _{-2}^{\nu }+2\theta _{\mu \nu }\alpha
_{-1}^{\mu }\alpha _{-3}^{\nu }\left\vert 0,k\right\rangle ;\left\vert
x\right\rangle =\theta _{\mu \nu }\alpha _{-1}^{\mu }\alpha _{-2}^{\nu
}\left\vert 0,k\right\rangle ,\text{ where }\theta _{\mu \nu }=-\theta _{\nu
\mu },  \tag{4.2}
\end{equation}

\begin{eqnarray}
L_{-1}\left\vert x\right\rangle &=&[2\theta _{\mu \nu }\alpha _{-2}^{\mu \nu
}+4\theta _{\mu \nu }\alpha _{-1}^{\mu }\alpha _{-3}^{\nu }+2(k_{\lambda
}\theta _{\mu \nu }+k_{(\lambda }\theta _{\mu \nu )})\alpha _{-1}^{\lambda
\mu }\alpha _{-2}^{\nu }+\frac{2}{3}k_{\lambda }k_{\beta }\theta _{\mu \nu
}\alpha _{-1}^{\mu \nu \lambda \beta }]\left\vert 0,k\right\rangle ;  \notag
\\
\left\vert x\right\rangle &=&[2\theta _{\mu \nu }\alpha _{-1}^{\mu }\alpha
_{-2}^{\nu }+\frac{2}{3}k_{\lambda }\theta _{\mu \nu }\alpha _{-1}^{\mu \nu
\lambda }]\left\vert 0,k\right\rangle ,  \TCItag{4.3}
\end{eqnarray}

\begin{eqnarray}
(L_{-2}+\frac{3}{2}L_{-1}^{2})\left\vert \widetilde{x}\right\rangle
&=&[3\theta _{\mu \nu }\alpha _{-2}^{\mu \nu }+8\theta _{\mu \nu }\alpha
_{-1}^{\mu }\alpha _{-3}^{\nu }+(k_{\lambda }\theta _{\mu \nu }+\frac{15}{2}%
k_{(\lambda }\theta _{\mu \nu )})\alpha _{-1}^{\lambda \mu }\alpha
_{-2}^{\nu }  \notag \\
&&+(\frac{1}{2}\eta _{\lambda \beta }\theta _{\mu \nu }+\frac{3}{2}%
k_{\lambda }k_{\beta }\theta _{\mu \nu })\alpha _{-1}^{\mu \nu \lambda \beta
}]\left\vert 0,k\right\rangle ;  \notag \\
\left\vert \widetilde{x}\right\rangle &=&\theta _{\mu \nu }\alpha _{-1}^{\mu
\nu }\left\vert 0,k\right\rangle  \TCItag{4.4}
\end{eqnarray}%
where $\alpha _{-1}^{\mu \nu }=\alpha _{-1}^{\mu }\alpha _{-1}^{\nu }$ etc.
There are two type I degenerate vector zero-norm states which can be
calculated as following

\begin{eqnarray}
\text{Ansatz} &:&\left\vert x\right\rangle =[a(\theta \cdot \alpha
_{-3})+b(k\cdot \alpha _{-2})(\theta \cdot \alpha _{-1})+c(k\cdot \alpha
_{-1})(\theta \cdot \alpha _{-2})  \notag \\
&&+d(\alpha _{-1}\cdot \alpha _{-1})(\theta \cdot \alpha _{-1})+f(k\cdot
\alpha _{-1})^{2}(\theta \cdot \alpha _{-1})]\left\vert 0,k\right\rangle . 
\TCItag{4.5}
\end{eqnarray}%
The $L_{1}$ and $L_{2}$ constraints of equation (2.1) give

\begin{equation}
a-2c=0,b+c+d-6f=0,3a-12b+28d-6f=0,  \tag{4.6}
\end{equation}%
which can be easily used to determine, for example, $a:b:c:d:f=26:5:13:0:3$
or $0:81:0:39:20.$ This gives two type I vector zero-norm states

\begin{eqnarray}
L_{-1}\left\vert x\right\rangle &=&[3a(\theta \cdot \alpha _{-4})+2b(k\cdot
\alpha _{-3})(\theta \cdot \alpha _{-1})+(2c+a)(k\cdot \alpha _{-1})(\theta
\cdot \alpha _{-3})  \notag \\
&&+(b+c)(k\cdot \alpha _{-2})(\theta \cdot \alpha _{-2})+(b+2f)(k\cdot
\alpha _{-1})(k\cdot \alpha _{-2})(\theta \cdot \alpha _{-1})  \notag \\
&&+2d(\alpha _{-2}\cdot \alpha _{-1})(\theta \cdot \alpha
_{-1})+(c+f)(k\cdot \alpha _{-1})^{2}(\theta \cdot \alpha _{-2})+d(\alpha
_{-1}\cdot \alpha _{-1})(\theta \cdot \alpha _{-2})  \notag \\
&&+d(k\cdot \alpha _{-1})(\alpha _{-1}\cdot \alpha _{-1})(\theta \cdot
\alpha _{-1})+f(k\cdot \alpha _{-1})^{3}(\theta \cdot \alpha
_{-1})]\left\vert 0,k\right\rangle .  \TCItag{4.7}
\end{eqnarray}%
The type II vector zero-norm state is

\begin{eqnarray}
(L_{-2}+\frac{3}{2}L_{-1}^{2})\left\vert \widetilde{x}\right\rangle
&=&[33(\theta \cdot \alpha _{-4})+4(k\cdot \alpha _{-3})(\theta \cdot \alpha
_{-1})+22(k\cdot \alpha _{-1})(\theta \cdot \alpha _{-3})  \notag \\
&&+\frac{21}{2}(k\cdot \alpha _{-2})(\theta \cdot \alpha _{-2})+\frac{11}{2}%
(k\cdot \alpha _{-1})(k\cdot \alpha _{-2})(\theta \cdot \alpha _{-1})  \notag
\\
&&+\frac{15}{2}(k\cdot \alpha _{-1})^{2}(\theta \cdot \alpha _{-2})+\frac{3}{%
2}(\alpha _{-1}\cdot \alpha _{-1})(\theta \cdot \alpha _{-2})  \notag \\
&&+\frac{1}{2}(k\cdot \alpha _{-1})(\alpha _{-1}\cdot \alpha _{-1})(\theta
\cdot \alpha _{-1})+\frac{3}{2}(k\cdot \alpha _{-1})^{3}(\theta \cdot \alpha
_{-1})]\left\vert 0,k\right\rangle ;  \notag \\
\left\vert \widetilde{x}\right\rangle &=&[3(\theta \cdot \alpha
_{-2})+(k\cdot \alpha _{-1})(\theta \cdot \alpha _{-1})]\left\vert
0,k\right\rangle .  \TCItag{4.8}
\end{eqnarray}%
The type I singlet zero-norm state was calculated to be \cite{5}

\begin{eqnarray}
\text{Ansatz} &:&\left\vert x\right\rangle =[a(k\cdot \alpha
_{-1})^{3}+b(k\cdot \alpha _{-1})(\alpha _{-1}\cdot \alpha _{-1})+c(k\cdot
\alpha _{-1})(k\cdot \alpha _{-2})  \notag \\
&&+d(\alpha _{-1}\cdot \alpha _{-2})+f(k\cdot \alpha _{-3})\left\vert
0,k\right\rangle .  \TCItag{4.9}
\end{eqnarray}%
The $L_{1}$ and $L_{2}$ constraints of equation (2.1) can be easily used to
determine $a:b:c:d:f=37:72:261:216:450.$ This gives the type I singlet
zero-norm state

\begin{eqnarray}
L_{-1}\left\vert x\right\rangle &=&[a(k\cdot \alpha _{-1})^{4}+b(k\cdot
\alpha _{-1})^{2}(\alpha _{-1}\cdot \alpha _{-1})+(2b+d)(k\cdot \alpha
_{-1})(\alpha _{-1}\cdot \alpha _{-2})  \notag \\
&&+(c+3a)(k\cdot \alpha _{-1})^{2}(k\cdot \alpha _{-2})+c(k\cdot \alpha
_{-2})^{2}+d(\alpha _{-2}\cdot \alpha _{-2})+b(k\cdot \alpha _{-2})(\alpha
_{-1}\cdot \alpha _{-1})  \notag \\
&&+(2c+f)(k\cdot \alpha _{-3})(k\cdot \alpha _{-1})+2d(\alpha _{-1}\cdot
\alpha _{-3})+3f(k\cdot \alpha _{-4})]\left\vert 0,k\right\rangle . 
\TCItag{4.10}
\end{eqnarray}%
Finally the type II singlet zero-norm state can be calculated to be

\begin{eqnarray}
(L_{-2}+\frac{3}{2}L_{-1}^{2})\left\vert \widetilde{x}\right\rangle
&=&[11a(k\cdot \alpha _{-4})+(6a+8c)(k\cdot \alpha _{-3})(k\cdot \alpha
_{-1})+8b(\alpha _{-1}\cdot \alpha _{-3})  \notag \\
&&+(\frac{5}{2}a+3c)(k\cdot \alpha _{-2})^{2}+(\frac{3}{2}a+\frac{17}{2}%
c)(k\cdot \alpha _{-1})^{2}(k\cdot \alpha _{-2})  \notag \\
&&+3b(\alpha _{-2}\cdot \alpha _{-2})+(\frac{5}{2}b+\frac{1}{2}a)(\alpha
_{-1}\cdot \alpha _{-1})(k\cdot \alpha _{-2})  \notag \\
&&+6b(k\cdot \alpha _{-1})(\alpha _{-2}\cdot \alpha _{-1})+(\frac{3}{2}b+%
\frac{1}{2}c)(k\cdot \alpha _{-1})^{2}(\alpha _{-1}\cdot \alpha _{-1}) 
\notag \\
&&+\frac{3}{2}c(k\cdot \alpha _{-1})^{4}+\frac{1}{2}b(\alpha _{-1}\cdot
\alpha _{-1})^{2}]\left\vert 0,k\right\rangle ;  \notag \\
\left\vert \widetilde{x}\right\rangle &=&[a(k\cdot \alpha _{-2})+b(\alpha
_{-1}\cdot \alpha _{-1})+c(k\cdot \alpha _{-1})^{2}]\left\vert
0,k\right\rangle  \TCItag{4.11}
\end{eqnarray}%
where $a:b:c=75:39:19.$ We are now ready to calculate the high-energy Ward
identities. The high-energy limit of stringy Ward identity corresponding to
eq.(4.1) are

\begin{equation}
\sqrt{6}(-\mathcal{T}_{LLLL}^{8\rightarrow 6}+3\mathcal{T}_{LLTT}^{6})+3(-%
\mathcal{T}_{LLL}^{6}+3\mathcal{T}_{LTT}^{4})=0,  \tag{4.12}
\end{equation}%
\begin{equation}
\sqrt{6}(-3\mathcal{T}_{LLLT}^{7\rightarrow 5}+\mathcal{T}_{LTTT}^{5})+3(-3%
\mathcal{T}_{LLT}^{5}+\mathcal{T}_{TTT}^{3})=0  \tag{4.13}
\end{equation}%
where $\mathcal{T}_{\mu \nu \lambda }$ is the amplitude corresponding to $%
\alpha _{-1}^{(\mu \nu }\alpha _{-2}^{\lambda )}.$ Eqs.(4.12) and (4.13)
correspond to $\theta ^{\mu \nu \lambda }=-e_{L}^{\mu }e_{L}^{\nu
}e_{L}^{\lambda }+3e_{(L}^{\mu }e_{T}^{\nu }$ $e_{T)}^{\lambda }$ and $%
\theta ^{\mu \nu \lambda }=-3e_{(L}^{\mu }e_{L}^{\nu }e_{T)}^{\lambda
}+e_{T}^{\mu }e_{T}^{\nu }e_{T}^{\lambda }$ respectively. Similarly eq.(4.2)
gives

\begin{equation}
\widetilde{\mathcal{T}}_{LL,T}^{5\rightarrow 3}+\sqrt{6}\widetilde{\mathcal{T%
}}_{[LT]}^{3}=0  \tag{4.14}
\end{equation}%
where$\widetilde{\mathcal{T}}_{\mu \nu }$ is the amplitude corresponding to $%
\alpha _{-1}^{\mu }\alpha _{-3}^{\nu }$ and $\widetilde{\mathcal{T}}_{\mu
\nu ,\lambda }$ is the amplitude corresponding to mixed symmetric part of $%
\alpha _{-1}^{\mu \nu }\alpha _{-2}^{\lambda },$ that is, first symmetrizing
w.r.t. $\mu \nu $ and then antisymmetrizing w.r.t. $\mu \lambda .$ This is
exactly the amplitude for the positive-norm mixed symmetric spn three state.
The type I symmetric spin two zero-norm state eq.(4.3) gives, in the
high-energy limit,

\begin{equation}
2(\mathcal{T}_{LLLL}^{8\rightarrow 6}-\mathcal{T}_{LLTT}^{6})+2\sqrt{6}[(%
\mathcal{T}_{LLL}^{6}-\mathcal{T}_{LTT}^{4})+\frac{1}{3}(\widetilde{\mathcal{%
T}}_{LL,P}^{6\rightarrow 4}+\widetilde{\mathcal{T}}_{LT,T}^{4})]+2(%
\widetilde{\mathcal{T}}_{(LL)}^{4}-\widetilde{\mathcal{T}}_{(TT)}^{2})+(%
\mathcal{T}_{LL}^{4}-\mathcal{T}_{TT}^{2})=0,  \tag{4.15}
\end{equation}

\begin{equation}
2\mathcal{T}_{LLLT}^{7\rightarrow 5}+\sqrt{6}[2\mathcal{T}_{LLT}^{5}+\frac{1%
}{3}\widetilde{\mathcal{T}}_{LL,T}^{5}]+2\widetilde{\mathcal{T}}_{(LT)}^{3}+%
\mathcal{T}_{LT}^{3}=0  \tag{4.16}
\end{equation}%
where $\mathcal{T}_{\mu \nu }$ is the amplitude corresponding to $\alpha
_{-2}^{\mu \nu }.$ The $E^{6}$ order of $\widetilde{\mathcal{T}}%
_{PL,L}^{6\rightarrow 4}$ in eq. (4.15) is forced to be zero in the
high-energy limit $(e_{P}=e_{L})$ due to the antisymmetric property of the
tensor $\widetilde{\mathcal{T}}_{\mu \nu ,\lambda }$ w.r.t.$\mu \lambda .$
It is important to note that in deriving eqs.(4.15) and (4.16), we have made
the following irreducible decomposition of the term

\begin{equation}
k_{\lambda }\theta _{\mu \nu }\alpha _{-1}^{\lambda \mu }\alpha _{-2}^{\nu
}=[\frac{1}{3}(k_{\lambda }\theta _{\mu \nu }+k_{\mu }\theta _{\nu \lambda
}+k_{\nu }\theta _{\lambda \mu })+\frac{1}{3}(k_{\lambda }\theta _{\mu \nu
}-k_{\nu }\theta _{\mu \lambda })]\alpha _{-1}^{\lambda \mu }\alpha
_{-2}^{\nu }  \tag{4.17}
\end{equation}%
in eq.(4.3). The first term with totally symmetric spin three index
corresponds to the gauge artifact of the positive-norm spin four state, and
the mixed symmetric tensor structure of the second term is exactly the same
as that of the positive-norm spin three state. In general, there are three
other possible mixed symmetric spin three terms, which do not appear in eq.
(4.17). This is a nontrivial consistent check of \ zero-norm states spectrum
in the OCFQ string. We will see similar mechanism happens in our later
calculation. The type II symmetric spin two zero-norm state eq.(4.4) gives,
in the high-energy limit,

\begin{equation}
\QATOP{9\mathcal{T}_{LLLL}^{8\rightarrow 6}-\frac{17}{2}\mathcal{T}%
_{LLTT}^{6}-\frac{1}{2}\mathcal{T}_{TTTT}^{4}+\frac{17}{2}\sqrt{6}[(\mathcal{%
T}_{LLL}^{6}-\mathcal{T}_{LTT}^{4})}{+\frac{2\sqrt{6}}{3}(\widetilde{%
\mathcal{T}}_{LL,P}^{6\rightarrow 4}+\widetilde{\mathcal{T}}_{LT,T}^{4})]+8(%
\widetilde{\mathcal{T}}_{(LL)}^{4}-\widetilde{\mathcal{T}}_{(TT)}^{2})+3(%
\mathcal{T}_{LL}^{4}-\mathcal{T}_{TT}^{2})=0,}  \tag{4.18}
\end{equation}

\begin{equation}
18\mathcal{T}_{LLLT}^{7\rightarrow 5}+\mathcal{T}_{LTTT}^{5}+\frac{4\sqrt{6}%
}{3}\widetilde{\mathcal{T}}_{LL,T}^{5}+17\sqrt{6}\mathcal{T}_{LLT}^{5}+16%
\widetilde{\mathcal{T}}_{(LT)}^{3}+6\mathcal{T}_{LT}^{3}=0.  \tag{4.19}
\end{equation}%
Two type I vector zero-norm states eq.(4.7) give, in the high-energy limit,

\begin{equation}
\QATOP{6\sqrt{6}f\mathcal{T}_{LLLL}^{8\rightarrow 6}+\sqrt{6}d\mathcal{T}%
_{LLTT}^{6}+6(b+c+3f)\mathcal{T}_{LLL}^{6}+3d\mathcal{T}_{LTT}^{4}}{+(4b-8c)%
\widetilde{\mathcal{T}}_{LP,P}^{6\rightarrow 4}+\sqrt{6}(2b+2c+a)\widetilde{%
\mathcal{T}}_{(LL)}^{4}+(2b-2c-a)\widetilde{\mathcal{T}}_{[LP]}^{4%
\rightarrow 2})+\sqrt{6}(b+c)\mathcal{T}_{LL}^{4}+3a\mathcal{T}_{L}^{2})=0,}
\tag{4.20}
\end{equation}

\begin{equation}
\QATOP{6\sqrt{6}f\mathcal{T}_{LLLT}^{7\rightarrow 5}+\sqrt{6}d\mathcal{T}%
_{LTTT}^{5}+6(b+c+3f)\mathcal{T}_{LLT}^{5}+3d\mathcal{T}_{TTT}^{3}}{-(4b-8c)%
\widetilde{\mathcal{T}}_{PP,T}^{5\rightarrow 3}+\sqrt{6}(2b+2c+a)\widetilde{%
\mathcal{T}}_{(LT)}^{3}+\sqrt{6}(2b-2c-a)\widetilde{\mathcal{T}}_{[TL]}^{3})+%
\sqrt{6}(b+c)\mathcal{T}_{LT}^{3}+3a\mathcal{T}_{T}^{1})=0}  \tag{4.21}
\end{equation}%
where $\mathcal{T}_{\mu }$ is the amplitude corresponding to $\alpha
_{-4}^{\mu }.$ Note that $\widetilde{\mathcal{T}}_{LP,P}^{6\rightarrow 4}$
in eq. (4.20) is identical to$\widetilde{\mathcal{T}}_{LL,P}^{6\rightarrow
4} $ in eqs.(4.15) and (4.18) in the high-energy limit. However, $\widetilde{%
\mathcal{T}}_{LP,P}^{2}$ and $\widetilde{\mathcal{T}}_{LL,P}^{2}$ can be
different. Also $\widetilde{\mathcal{T}}_{PP,T}^{5}$ in eq.(4.21) is zero
since it equals to $\widetilde{\mathcal{T}}_{LL,T}^{5}$ in eq.(4.14), which
is zero, in the high-energy limit. However, $\widetilde{\mathcal{T}}%
_{PP,T}^{3}$ and $\widetilde{\mathcal{T}}_{LL,T}^{3}$ can be different. In
deriving eqs.(4.20) and (4.21), in addition to (4.17), one needs another
projection formula

\begin{equation}
k_{\lambda }k_{\mu }\theta _{\nu }\alpha _{-1}^{\lambda \mu }\alpha
_{-2}^{\nu }=[\frac{1}{3}(k_{\lambda }k_{\mu }\theta _{\nu }+k_{\mu }k_{\nu
}\theta _{\lambda }+k_{\nu }k_{\lambda }\theta _{\mu })+\frac{2}{3}%
(k_{\lambda }k_{\mu }\theta _{\nu }-k_{\nu }k_{\mu }\theta _{\lambda
})]\alpha _{-1}^{\lambda \mu }\alpha _{-2}^{\nu }.  \tag{4.22}
\end{equation}%
Again, the first term of eq.(4.22) with totally symmetric spin three index
corresponds to the gauge artifact of the positive-norm spin four state, and
the mixed symmetric tensor structure of the second term is exactly the same
as that of the positive-norm spin three state. This is another consistent
check of \ zero-norm states spectrum in the OCFQ string. In the following,
we will use eqs.(4.17) and (4.22) whenever they are needed. Type II vector
zero-norm state eq.(4.6) gives, in the high-energy limit,

\begin{equation}
\QATOP{9\sqrt{6}\mathcal{T}_{LLLL}^{8\rightarrow 6}+\frac{\sqrt{6}}{2}%
\mathcal{T}_{LLTT}^{6}+78\mathcal{T}_{LLL}^{6}+\frac{3}{2}\mathcal{T}%
_{LTT}^{4}+2\widetilde{\mathcal{T}}_{LT,T}^{4}}{+38\widetilde{\mathcal{T}}%
_{LP,P}^{6\rightarrow 4}+26\sqrt{6}\widetilde{\mathcal{T}}_{(LL)}^{4}-18%
\widetilde{\mathcal{T}}_{[LP]}^{4\rightarrow 2}+\frac{21}{2}\sqrt{6}\mathcal{%
T}_{LL}^{4}+33\mathcal{T}_{L}^{2}=0,}  \tag{4.23}
\end{equation}

\begin{equation}
\QATOP{9\sqrt{6}\mathcal{T}_{LLLT}^{7\rightarrow 5}+\frac{\sqrt{6}}{2}%
\mathcal{T}_{LTTT}^{5}+78\mathcal{T}_{LLT}^{5}+\frac{3}{2}\mathcal{T}%
_{TTT}^{3}}{+38\widetilde{\mathcal{T}}_{TL,L}^{5}+26\sqrt{6}\widetilde{%
\mathcal{T}}_{(LT)}^{3}-18\widetilde{\mathcal{T}}_{[TL]}^{3}+\frac{21}{2}%
\sqrt{6}\mathcal{T}_{LT}^{3}+33\mathcal{T}_{T}^{1}=0.}  \tag{4.24}
\end{equation}%
Note that $\widetilde{\mathcal{T}}_{TL,L}^{5}$ in eq.(4.23) is identical to $%
\widetilde{\mathcal{T}}_{TP,P}^{5}$ in eq. (4.21) in the high-energy limit.
Finally, type I and type II singlet zero-norm states give, in the
high-energy limit,

\begin{equation}
\QATOP{74\mathcal{T}_{LLLL}^{8\rightarrow 6}+24\mathcal{T}_{LLTT}^{6}+124%
\sqrt{6}\mathcal{T}_{LLL}^{6}+24\sqrt{6}\mathcal{T}_{LTT}^{4}-8\sqrt{6}%
\widetilde{\mathcal{T}}_{LT,T}^{4}}{+324\widetilde{\mathcal{T}}_{(LL)}^{4}+87%
\mathcal{T}_{LL}^{4}=0,}  \tag{4.25}
\end{equation}

\begin{equation}
\QATOP{342\mathcal{T}_{LLLL}^{8\rightarrow 6}+136\mathcal{T}_{LLTT}^{6}+%
\frac{13}{2}\mathcal{T}_{TTTT}^{4}+548\sqrt{6}\mathcal{T}_{LLL}^{6}+123\sqrt{%
6}\mathcal{T}_{LTT}^{4}+8\sqrt{6}\widetilde{\mathcal{T}}_{LT,T}^{4}}{+1204%
\widetilde{\mathcal{T}}_{(LL)}^{4}+489\mathcal{T}_{LL}^{4}=0.}  \tag{4.26}
\end{equation}%
This completes the calculation of high-energy Ward identities. It is easy to
count the high-energy amplitudes for each tensor. For $\mathcal{T}_{\mu \nu
\lambda \gamma }$, one has $\mathcal{T}_{LLLL},\mathcal{T}_{LLLT},\mathcal{T}%
_{LLTT},\mathcal{T}_{LTTT}$ and $\mathcal{T}_{TTTT}$. For $\mathcal{T}_{\mu
\nu \lambda }$, one has $\mathcal{T}_{LLL},\mathcal{T}_{LLT},\mathcal{T}%
_{LTT}$ and $\mathcal{T}_{TTT}.$ For $\widetilde{\mathcal{T}}_{\mu \nu
,\lambda },$ one has $\widetilde{\mathcal{T}}_{LL,T}$ and $\widetilde{%
\mathcal{T}}_{LT,T}.$ For $\mathcal{T}_{\mu \nu ,\text{ }}$one has $\mathcal{%
T}_{LL},\mathcal{T}_{LT}$ and $\mathcal{T}_{TT}.$ For $\widetilde{\mathcal{T}%
}_{\mu \nu ,}$ one has $\widetilde{\mathcal{T}}_{LL,}\widetilde{\mathcal{T}}%
_{(LT),}\widetilde{\mathcal{T}}_{[LT]}$ and $\widetilde{\mathcal{T}}_{TT}.$
For $\mathcal{T}_{\mu ,\text{ }}$one has $\mathcal{T}_{L}$ and $\mathcal{T}%
_{T}.$ It is very important to note that in the $E^{4}$ order, one gets one
more amplitude $\widetilde{\mathcal{T}}_{LP,P}^{4}$, and in the $E^{3}$
order, one gets another amplitude $\widetilde{\mathcal{T}}_{PP,T}^{3}$
described after eq.(4.21). It can be checked by eqs.(4.12)-(4.26) that all
the amplitudes of orders $E^{8}$ $E^{7}$ $E^{6}$ and $E^{5}$ are zero. So
the real leading order amplitudes correspond to $E^{4}$, which appear in
eqs.(4.12), (4.15), (4.18), (4.20), (4.23), (4.25) and (4.26). Note that
there are two equations for (4.20). We thus end up with 8 equations and 9
amplitudes. A calculation shows that

\begin{eqnarray}
\mathcal{T}_{TTTT}^{4} &:&\mathcal{T}_{TTLL}^{4}:\mathcal{T}_{LLLL}^{4}:%
\mathcal{T}_{TTL}^{4}:\mathcal{T}_{LLL}^{4}:\widetilde{\mathcal{T}}%
_{LT,T}^{4}:\widetilde{\mathcal{T}}_{LP,P}^{4}:\mathcal{T}_{LL}^{4}:%
\widetilde{\mathcal{T}}_{LL}^{4}=  \notag \\
16 &:&\frac{4}{3}:\frac{1}{3}:-\frac{4\sqrt{6}}{9}:-\frac{\sqrt{6}}{9}:-%
\frac{2\sqrt{6}}{3}:0:\frac{2}{3}:0.  \TCItag{4.27}
\end{eqnarray}%
Note that these proportionality constants are again, as conjectured by
Gross, independent of the scattering angle $\phi _{CM}$ and the loop order $%
\chi $ of string perturbation theory. They are also independent of particles
chosen for vertex $v_{1,3,4}$. The subleading order amplitudes corresponding
to $E^{3}$ appear in eqs.(4.13), (4.14), (4.16), (4.19), (4.21) and (4.24).
Note that there are two equations for (4.21). One has 7 equations with 9
amplitudes, $\mathcal{T}_{TTTL}^{3},\mathcal{T}_{TLLL}^{3},\mathcal{T}%
_{TLL}^{3},\mathcal{T}_{TTT}^{3},\widetilde{\mathcal{T}}_{TL,L}^{3},%
\widetilde{\mathcal{T}}_{PP,T}^{3},\mathcal{T}_{LT}^{3},\widetilde{\mathcal{T%
}}_{(LT)}^{3}$ and $\widetilde{\mathcal{T}}_{[LT]}^{3}.$ Presumably, they
are not proportional to each other or the proportional coefficients do
depend on the scattering angle $\phi _{CM}$. Our calculation here is again
purely algebraic \textit{without any integration} and is independent of
saddle point calculation in \cite{3,4,10}. It is important to note that our
result in eq.(4.27) is gauge invariant. On the other hand, the result
obtained in \cite{10} with $\mathcal{T}_{TTTT}^{4}\propto \widetilde{%
\mathcal{T}}_{LT,T}^{4}\propto \mathcal{T}_{LL}^{4},$ and $\mathcal{T}%
_{TTLL}^{4}=\mathcal{T}_{LLLL}^{4}=\mathcal{T}_{TTL}^{4}=\mathcal{T}%
_{LLL}^{4}=\widetilde{\mathcal{T}}_{LP,P}^{4}=\widetilde{\mathcal{T}}%
_{LL}^{4}=0$ in the leading order energy is, on the contrary, \textit{not}
gauge invariant. In fact, with only three non-zero amplitudes, it would be
very difficult to satisfy all 8 equations. The situation gets even worse if
one goes to higher mass level where number of zero-norm states, or
constraint equations, increases much faster than that of positive-norm
states \cite{5}. To further justify our result, we give a sample calculation
in the following.

Since the proportionality constants in eq.(4.27) are independent of
particles chosen for vertex $v_{1,3,4}.$ For simplicity, we will choose them
to be tachyons. For the string-tree level $\chi $=1, with one tensor $v_{2}$
and three tachyons $v_{1,3,4}$, all scattering amplitudes for mass level $%
M^{2}$ $=6$ were explicitly calculated in \cite{14}. They are

\begin{eqnarray}
\mathcal{T}^{\mu \nu \alpha \beta } &=&\tint
\tprod_{i=1}^{4}dx_{i}<e^{ik_{1}X}\partial X^{\mu }\partial X^{\nu }\partial
X^{\alpha }\partial X^{\beta }e^{ik_{2}X}e^{ik_{3}X}e^{ik_{4}X}>  \notag \\
&=&\frac{\Gamma (-\frac{s}{2}-1)\Gamma (-\frac{t}{2}-1)}{\Gamma (\frac{u}{2}%
+2)}[(\frac{s^{2}}{4}-s)(\frac{s^{2}}{4}-1)k_{3}^{\mu }k_{3}^{\nu
}k_{3}^{\alpha }k_{3}^{\beta }  \notag \\
&&-t(\frac{t^{2}}{4}-1)(s+2)k_{1}^{(\mu }k_{1}^{\nu }k_{1}^{\alpha
}k_{3}^{\beta )}+\frac{3st}{2}(\frac{s}{2}+1)(\frac{t}{2}+1)k_{1}^{(\mu
}k_{1}^{\nu }k_{3}^{\alpha }k_{3}^{\beta )}  \notag \\
&&-s(\frac{s^{2}}{4}-1)(t+2)k_{1}^{(\mu }k_{3}^{\nu }k_{3}^{\alpha
}k_{3}^{\beta )}+(\frac{t^{2}}{4}-t)(\frac{t^{2}}{4}-1)k_{1}^{\mu
}k_{1}^{\nu }k_{1}^{\alpha }k_{1}^{\beta }],  \TCItag{4.28}
\end{eqnarray}

\begin{eqnarray}
\mathcal{T}^{\mu \nu \lambda } &=&\tint
\tprod_{i=1}^{4}dx_{i}<e^{ik_{1}X}\partial X^{\mu }\partial X^{\nu }\partial
^{2}X^{\lambda }e^{ik_{2}X}e^{ik_{3}X}e^{ik_{4}X}>  \notag \\
&=&\frac{\Gamma (-\frac{s}{2}-1)\Gamma (-\frac{t}{2}-1)}{\Gamma (\frac{u}{2}%
+2)}[-(\frac{s^{2}}{4}-s)(\frac{s^{2}}{4}-1)k_{3}^{\mu }k_{3}^{\nu
}k_{3}^{\lambda }  \notag \\
&&+t(\frac{t^{2}}{4}-1)(\frac{s}{2}+1)k_{1}^{\lambda }k_{1}^{(\mu
}k_{3}^{\nu )}-\frac{st}{4}(\frac{s}{2}+1)(\frac{t}{2}+1)(k_{1}^{\mu
}k_{1}^{\nu }k_{3}^{\lambda }+k_{3}^{\mu }k_{3}^{\nu }k_{1}^{\lambda }) 
\notag \\
&&+s(\frac{s^{2}}{4}-1)(\frac{t}{2}+1)k_{3}^{\lambda }k_{1}^{(\mu
}k_{3}^{\nu )}-(\frac{t^{2}}{4}-t)(\frac{t^{2}}{4}-1)k_{1}^{\mu }k_{1}^{\nu
}k_{1}^{\lambda }],  \TCItag{4.29}
\end{eqnarray}

\begin{eqnarray}
\mathcal{T}^{\mu \nu } &=&\tint \tprod_{i=1}^{4}dx_{i}<e^{ik_{1}X}\partial
^{2}X^{\mu }\partial ^{2}X^{\nu }e^{ik_{2}X}e^{ik_{3}X}e^{ik_{4}X}>  \notag
\\
&=&\frac{\Gamma (-\frac{s}{2}-1)\Gamma (-\frac{t}{2}-1)}{\Gamma (\frac{u}{2}%
+2)}[(\frac{s^{2}}{4}-s)(\frac{s^{2}}{4}-1)k_{3}^{\mu }k_{3}^{\nu }  \notag
\\
&&+\frac{st}{2}(\frac{s}{2}+1)(\frac{t}{2}+1)k_{1}^{(\mu }k_{3}^{\nu )}+(%
\frac{t^{2}}{4}-t)(\frac{t^{2}}{4}-1)k_{1}^{\mu }k_{1}^{\nu }], 
\TCItag{4.30}
\end{eqnarray}

\begin{eqnarray}
\widetilde{\mathcal{T}}^{\mu \nu } &=&\frac{1}{2}\tint
\tprod_{i=1}^{4}dx_{i}<e^{ik_{1}X}\partial X^{\mu }\partial ^{3}X^{\nu
}e^{ik_{2}X}e^{ik_{3}X}e^{ik_{4}X}>  \notag \\
&=&\frac{\Gamma (-\frac{s}{2}-1)\Gamma (-\frac{t}{2}-1)}{\Gamma (\frac{u}{2}%
+2)}[(\frac{s^{2}}{4}-s)(\frac{s^{2}}{4}-1)k_{3}^{\mu }k_{3}^{\nu }-\frac{s}{%
2}(\frac{s^{2}}{4}-1)(\frac{t}{2}+1)k_{1}^{\mu }k_{3}^{\nu }  \notag \\
&&-\frac{t}{2}(\frac{t^{2}}{4}-1)(\frac{s}{2}+1)k_{3}^{\mu }k_{1}^{\nu }+(%
\frac{t^{2}}{4}-t)(\frac{t^{2}}{4}-1)k_{1}^{\mu }k_{1}^{\nu }], 
\TCItag{4.31}
\end{eqnarray}

\begin{eqnarray}
\mathcal{T}^{\mu } &=&\frac{1}{6}\tint
\tprod_{i=1}^{4}dx_{i}<e^{ik_{1}X}\partial ^{4}X^{\mu
}e^{ik_{2}X}e^{ik_{3}X}e^{ik_{4}X}>  \notag \\
&=&\frac{\Gamma (-\frac{s}{2}-1)\Gamma (-\frac{t}{2}-1)}{\Gamma (\frac{u}{2}%
+2)}[-(\frac{s^{2}}{4}-s)(\frac{s^{2}}{4}-1)k_{3}^{\mu }-(\frac{t^{2}}{4}-t)(%
\frac{t^{2}}{4}-1)k_{1}^{\mu }].  \TCItag{4.32}
\end{eqnarray}%
We can now explicitly calculate all amplitudes in eq.(4.27). After a lengthy
algebra, we have justified eq.(4.27) with $\mathcal{T}_{TTTT}^{4}=16E^{12}%
\sin ^{4}\phi _{CM}\mathcal{T}(4)$ in the high-energy limit. We have also
checked that $\mathcal{T}_{LLLL}^{8}=\mathcal{T}_{LLLL}^{6}=\mathcal{T}%
_{LLL}^{6}=\mathcal{T}_{TTLL}^{6}=\widetilde{\mathcal{T}}_{LP,P}^{6}=%
\widetilde{\mathcal{T}}_{LL,P}^{6}=0$ and $\widetilde{\mathcal{T}}%
_{LP,P}^{4} $ =$\widetilde{\mathcal{T}}_{LL,P}^{4}$ as claimed above. The
calculation of $\mathcal{T}_{LLLL},$for example, gives

\begin{eqnarray}
\mathcal{T}_{LLLL} &=&\frac{\Gamma (-\frac{s}{2}-1)\Gamma (-\frac{t}{2}-1)}{%
\Gamma (\frac{u}{2}+2)}[(\frac{s^{2}}{4}-s)(\frac{s^{2}}{4}%
-1)(e_{L}.k_{3})^{4}  \notag \\
&&-s(\frac{s^{2}}{4}-1)(t+2)(e_{L}.k_{3})^{3}(e_{L}.k_{1})+\frac{3st}{2}(%
\frac{s}{2}+1)(\frac{t}{2}+1)(e_{L}.k_{3})^{2}(e_{L}.k_{1})^{2}  \notag \\
&&-t(\frac{t^{2}}{4}-1)(s+2)(e_{L}.k_{3})(e_{L}.k_{1})^{3}+(\frac{t^{2}}{4}%
-t)(\frac{t^{2}}{4}-1)(e_{L}.k_{1})^{4}].  \TCItag{4.33}
\end{eqnarray}%
By using eqs.(3.6), (3.9), (3.11) and (3.12) and after a lengthy algebra, we
find that the contributions of orders $E^{16}$ and $E^{14}$ of $\mathcal{T}%
_{LLLL}$ are zero. The leading order $E^{12}$ term gives $\mathcal{T}%
_{LLLL}^{4}=\frac{1}{3}\sin ^{4}\phi _{CM}E^{12}\mathcal{T}(4)$ as expected
from eq. (4.27). Similar calculations can be done for other 8 amplitudes,
and eq. (4.27) is justified after a long calculation. Finally, by eqs.(3.1),
(3.7), (3.10) and (4.28), it is easy to deduce in general that 
\begin{equation}
\mathcal{T}_{n}^{TT...}=[(-2)^{n}E^{3n}\sin ^{n}\phi _{CM}]\mathcal{T}(n), 
\tag{4.34}
\end{equation}%
where $n$ is the number of $T$ and $\mathcal{T}(0)$ is the high energy four
tachyons amplitude. As a result, all high-energy string scattering
amplitudes can be expressed in terms of those of tachyons.

\section{A comparision with saddle point calculation}

To compare our results with Ref \cite{10}, we briefly review the works in 
\cite{3,4,10}. In Ref \cite{10}, it was shown that the high-energy, fixed
angle scattering amplitudes of oriented open strings can be obtained from
those of closed strings calculated by Gross and Mende \cite{3} by using the
reflection principle. First, from eq. (2.3), one notes that the high-energy
limit $\alpha ^{\prime }\rightarrow \infty $ is equivalent to the
semi-classical limit of first-quantized string theory. In this limit, the
closed string G-loop scattering amplitudes is dominated by a saddle point in
the moduli space $\overrightarrow{m}$. For the oriented open string
amplitudes, the saddle point configuration can be constructed from an
associated configuration of the closed string via reflection principle. It
was also found that the Euler number $\chi $ of the oriented open string
saddle is always $%
\chi =1-G%
$, where G is the genus of the associated closed string saddle. Thus the
integral in eq. (2.3) is dominated in the $\alpha ^{\prime }\rightarrow
\infty $ limit by an associated G-loop closed string saddle point in $X^{\mu
}$,$\widehat{\overrightarrow{m}_{i}}$ and $\widehat{\xi _{i}}$. The closed
string classical trajectory at G-loop order was found to behave at the
saddle point as \cite{3}

\begin{equation}
X_{c1}^{\mu }(z)=\frac{i}{1+G}\overset{4}{\underset{i=1}{\sum }}k_{i}^{\mu
}\ln \left\vert z-a_{i}\right\vert +O(\frac{1}{\alpha ^{\prime }}), 
\tag{5.1}
\end{equation}%
which leads to the $\chi $-th order open string four-tachyon amplitude

\begin{equation}
\mathcal{T}_{\chi }\approx g_{c}^{2-\chi }\exp (-\alpha ^{\prime }\frac{s\ln
s+t\ln t+u\ln u}{2(2-\chi )}).  \tag{5.2}
\end{equation}%
Eq. (5.2) reproduces the very soft exponential decay e$^{-\alpha ^{\prime
}s} $ of the well-known string-tree $\chi $=1 amplitude. The exponent of
eq.(5.2) can be thought of as the electrostatic energy $E_{G}$ of
two-dimensional Minkowski charges $k_{i}$ placed at $a_{i}$ on a Riemann
surface of genus $G$. One can use the $SL(2,C)$ invariance of the saddle to
fix 3 of the 4 points $a_{i}$, then the only modulus is the cross ratio $%
\lambda =\frac{(a_{1}-a_{3})(a_{2}-a_{4})}{(a_{1}-a_{2})(a_{3}-a_{4})}$,
which takes the value $\lambda =\widehat{\lambda }\approx -\frac{t}{s}%
\approx \sin ^{2}\frac{\phi _{CM}}{2}$ to extremize $E_{G}$ if we neglect
the mass of the tachyons in the high-energy limit. For excited string
states, it was found that only polarizations in the plane of scattering will
contribute to the amplitude at high energy. To leading order in the energy $%
E $, the products of e$_{T}$ and e$_{L}$ with $\partial ^{n}X$ are given by 
\cite{10}

\begin{equation}
e_{T}\cdot \partial ^{n}X\sim i(-)^{n}\frac{(n-1)!}{\lambda ^{n}}E\sin \phi
_{CM},n>0;  \tag{5.3}
\end{equation}

\begin{equation}
e_{L}\cdot \partial ^{n}X\sim i(-)^{(n-1)}\frac{(n-1)!}{\lambda ^{n}}\frac{%
E^{2}\sin ^{2}\phi _{CM}}{2m_{2}}\overset{n-2}{\underset{l=0}{\sum }\lambda
^{l}},n>1;  \tag{5.4}
\end{equation}%
\begin{equation}
e_{L}\cdot \partial ^{n}X\sim 0,n=1,\text{\ \ \ \ \ \ \ \ \ \ \ \ \ \ \ \ \
\ \ \ \ \ \ \ \ \ \ \ \ }  \tag{5.5}
\end{equation}%
where $m_{2}$ is the mass of the particle. Now, we would like to point out
that naive uses of eqs.(5.3) to (5.5) will miss some high-energy amplitudes
and will give, for example, a wrong result $\mathcal{T}_{LLT}^{3}=0$ \cite%
{10} since $e_{L}\cdot \partial X\sim 0.$This is inconsistent with our
result eq.(2.19) or eq.(3.14). The missing terms can be seen as following.
We will use the $M^{2}=4$ string-tree $\chi =1$ amplitude $\mathcal{T}_{LLT} 
$ to illustrate our point. Let's first use the path integral calculation

\begin{equation}
\mathcal{T}_{LLT}=\tint \tprod_{i=1}^{4}dx_{i}<e^{ik_{1}X}e_{L}\cdot
\partial Xe_{L}\cdot \partial Xe_{T}\cdot \partial
Xe^{ik_{2}X}e^{ik_{3}X}e^{ik_{4}X}>,  \tag{5.6}
\end{equation}%
which is similar to the calculation of moments of the Gaussian integral

\begin{equation}
\sqrt{\frac{a}{2\pi }}\dint\limits_{-\infty }^{\infty }dxx^{n}e^{-\frac{a}{2}%
x^{2}+bx}=\frac{\partial ^{n}}{\partial b^{n}}\sqrt{\frac{a}{2\pi }}%
\dint\limits_{-\infty }^{\infty }dxe^{-\frac{a}{2}x^{2}+bx}.  \tag{5.7}
\end{equation}%
For $n=1$, the value obtained by eq.(5.7) is $\frac{b}{a}e^{\frac{b^{2}}{2a}%
}=xe^{-\frac{a}{2}x^{2}+bx}\mid _{x=\frac{b}{a}}$where $\frac{b}{a}$ is
exactly the saddle point of the Gaussian integrand. For $n=2$, however, the
value obtained by eq.(5.7) is $(\frac{b}{a})^{2}e^{\frac{b^{2}}{2a}}+\frac{1%
}{a}e^{\frac{b^{2}}{2a}}=x^{2}e^{-\frac{a}{2}x^{2}+bx}\mid _{x=\frac{b}{a}}+%
\frac{1}{a}e^{\frac{b^{2}}{2a}}.$ It is this extra $\frac{1}{a}e^{\frac{b^{2}%
}{2a}}$ term that was missing in the argument of section 6 of Ref \cite{10}.
Similar situations happen for $n\geqslant 3$ and even more terms were
missed. The argument can be easily generalized to $\overrightarrow{b}$ $\in
R^{3}$ in the space of helixity decomposition$.$ Eq.(5.6) corresponds to the
case of $n=3$. It can be checked that some terms with the same energy order
as $\mathcal{T}_{TTT}$ survive in the calculation of eq.(5.6). They will be
missing if one misuses eqs.(5.3) to (5.5). Similar wrong calculations will
suppress many other should be non-zero high-energy amplitudes at mass level $%
M^{2}=6$ stated after eq.(4.27). Another way to calculate eq.(5.6) is to use
Wick theorem. Again, naive uses of eqs.(5.3) to (5.5) will miss some
high-energy amplitudes which correspond to, for example, the contraction of $%
e^{ik_{1}X}$ with $(e_{L}\cdot \partial X)(e_{L}\cdot \partial X).$ We
stress here that eqs.(5.1) to (5.5) are still valid as they stand.

\section{Conclusion}

We have shown that the physical origin of high-energy symmetries and the
proportionality constants in eqs. (2.19) and (4.27) are from zero-norm
states in the OCFQ spectrum. Other related approaches of high-energy stringy
symmetries can be found in \cite{15}. The most challenging problem remained
is the calculation of algebraic structure of these stringy symmetries
derived from the complete zero-norm state solutions of eqs. (2.1) and (2.2)
with arbitrarily high spins. Presumably, it is a complicated 26D
generalization of $\omega _{\infty }$ of the simpler toy 2D string model 
\cite{8}. Our calculation in eqs.(2.19) and (4.27) are, similar to the toy
2D string, purely algebraic without any integration which signal the
powerfulness of zero-norm states and symmetries they imply. The results
presented in this paper can be served as consistent checks of saddle point
calculations \cite{3} and as the realization of high-energy symmetries \cite%
{4} of string theory. The simple idea of massive gauge invariance of our
calculation corrects the inconsistent high-energy calculation in section 6
of Ref \cite{10}.

\section{Acknowledgments}

We would like to thank Physics Department of National Taiwan University for
the hospitality and computer facilities. This work is supported by grants of
NSC of Taiwan.

\end{document}